\documentstyle[12pt]{article}
\newcommand{\delslash}{\not \! \partial} 
 
\begin{document}

\begin{flushright}{UT-03-32\\TUW-03-29\\YITP-SB-03-49}
\end{flushright}
\vskip 0.5 truecm

\begin{center}
{\Large{\bf On the nature of the anomalies in the\\[3pt]
supersymmetric kink}} 
\end{center} 
\vskip .5 truecm 
\centerline{\bf Kazuo Fujikawa\footnote{
         fujikawa@phys.s.u-tokyo.ac.jp} }
\vskip .4 truecm
\centerline {\it Department of Physics,University of Tokyo} 
\centerline {\it Bunkyo-ku,Tokyo 113,Japan} 
\vskip 0.5 truecm 
\centerline{\bf Anton Rebhan\footnote{  
rebhana@hep.itp.tuwien.ac.at}} 
\vskip .4 truecm 
\centerline{\it Institut f\"ur Theoretische Physik, 
Technische Universit\"at Wien} 
\centerline{\it Wiedner Hauptstr. 8--10, 1040 Vienna, Austria} 
\vskip 0.5 truecm 
\centerline{\bf Peter van Nieuwenhuizen\footnote{ 
 vannieu@insti.physics.sunysb.edu}} 
\vskip .4 truecm 
\centerline{\it C.N. Yang Institute for Theoretical Physics} 
\centerline{\it State University of New York, Stony Brook, NY 
11794-3840, USA} 
\vskip 0.5 truecm

\makeatletter
\@addtoreset{equation}{section} 
\def\theequation{\thesection.\arabic{equation}}
\makeatother

\setcounter{footnote}{0}

\begin{abstract}
We discuss the 
possibility to absorb
all anomalies in the supersymmetry algebra of the $N=(1,1)$
Wess-Zumino model in $d=1+1$ by a local counter term. This 
counter term corresponds to 
the change of the vacuum parameter $v_{0}^{2}$ in the model 
and the transition to an unconventional but 
admissible renormalization scheme. 
It does not modify the physical consequences such 
as BPS saturation,
and thus the situation is  rather different from 
 gauge theory where local counter terms are required to
absorb spurious gauge anomalies.
\end{abstract}


\section{Introduction}

In the past few years it has been established that the 
one-loop quantum corrections to the mass and  central charge 
of the supersymmetric kink in $d=1+1$ 
dimensions, which were studied long ago in~\cite{R3,Dashen:1974cj,kaul}, are 
non-vanishing 
\cite{Rebhan:1997iv,Nastase:1998sy,Goldhaber:2000ab,Rebhan:2002uk} 
and equal \cite{Graham:1998qq, Shifman:1998zy,
losev, Bordag:2002dg,
GLN, Rebhan:2002yw}.
The corrections to the mass 
involve various subtleties depending on the regularization methods,
but the corresponding corrections to the central charge 
posed initially a particular problem. The central charge is the integral of 
a total derivative and 
it seemed not to receive
any corrections due to the presence of the kink 
soliton (in a minimal renormalization scheme
with only mass renormalization such that
tadpoles vanish)~\cite{Rebhan:1997iv,Nastase:1998sy}.
On the other hand, although 
saturation of the BPS bound seemed likely~\cite{R3},
no formal proof was available at that time 
and it was conjectured 
that 
a {new} anomaly is responsible for the discrepancy of the quantum 
corrections~\cite{Nastase:1998sy}. Other 
authors~\cite{Shifman:1998zy,
losev, Bordag:2002dg, 
GLN, Rebhan:2002yw} have since then 
established the existence of a
central charge anomaly which
is connected to the trace anomaly through supersymmetry and which
is responsible for the saturation of the BPS bound at the quantum
level.\footnote{ It should be noted, 
however, that \cite{Graham:1998qq} did not find an anomalous 
contribution to the central charge operator. According to  
\cite{Rebhan:2002yw}, this is due to the fact that 
\cite{Graham:1998qq} employed a consistent regularization only when 
considering the kink mass, but not when relating the central 
charge operator to the Hamiltonian in a formal (unregularized) 
manner.}

Recently,  a superspace path integral 
analysis was made~\cite{fvn} according to which all anomalies 
reside in one super Jacobian~\cite{fuji1,fuji2,konishi,shizuya}.
 The simplicity of the 
results in the  superspace formulation revealed that the anomaly
  in superspace (and thus also 
all the anomalies in $x$-space) can formally be removed 
by adding an order-$\hbar$ counter term to the action. In  
renormalizable gauge 
theories in 4-dimensions one {\em must} in such cases add such a 
counter term 
to the action, in order that higher-loop renormalizability 
be preserved. We argue, however, that in  the present 
super-renormalizable model 
the situation is more subtle. In addition to anomalies, there is
 spontaneous $Z_{2}$ symmetry breaking and {\em explicit} conformal 
symmetry breaking present, and depending on 
whether or not one adds a counter term, the various quantum 
currents which one can define change. In fact, we derive below 
relations between conserved currents with the counter term 
present and non-conserved currents without the counter term.  
The various relations become quite clear if one uses path 
integrals, and explain why after adding the counter term one 
still has the same physical content while keeping the essence 
of the trace and central charge anomalies. 

\section{Previous results}

The $N=(1,1)$ Wess-Zumino model in $d=1+1$ dimensions is 
defined by 
\begin{eqnarray}\label{Lagr}
&&\int dx d^{2}\theta {\cal L}(x,\theta)=\int dxd^{2}\theta 
[\frac{1}{4}\bar{D}\phi(x,\theta)D\phi(x,\theta)
+\frac{1}{3}g\phi^{3}(x,\theta)
-gv^{2}_{0}\phi(x,\theta)]\nonumber\\
&&=\int dx\{\frac{1}{2}[FF 
-\partial^{\mu}\varphi\partial_{\mu}\varphi
-\bar{\psi}\gamma^{\mu}\partial_{\mu}\psi]
+gF\varphi^{2}-g(\bar{\psi}\psi)\varphi
-gv^{2}_{0}F\}.
\end{eqnarray}
The coupling constant $g$ has mass dimension one and $v_{0}$ is 
dimensionless.
The superfield $\phi$ is defined by
\begin{equation}
\phi(x,\theta)=\varphi(x)+\bar{\theta}\psi(x)
+\frac{1}{2}\bar{\theta}\theta F(x)
\end{equation}
where $\theta^{\alpha}$ is a Grassmann number, and 
$\theta^{\alpha}$ and $\psi^{\alpha}(x)$ are two-component 
Majorana spinors; 
$\varphi(x)$ is a real 
scalar field, and $F(x)$ is a real auxiliary field. 
We define $\bar{\theta}=\theta^{T}C$ with $C$ the charge 
conjugation matrix, and the inner product for spinors is defined
  by 
\begin{equation}
\bar{\theta}\theta\equiv \theta^{T}C\theta 
= \theta^{\alpha} C_{\alpha\beta}\theta^{\beta}\equiv
\bar{\theta}_{\beta}\theta^{\beta}
\end{equation}
with the Dirac  matrix convention
\begin{equation}
\gamma^{0}=-\gamma_{0}=-i\tau^{2}, \ \ \ \gamma^{1}=\gamma_{1}
= \tau^{3},\ \ \ 
C=\tau^{2}, \ \ \ \gamma_{5}=\gamma^{0}\gamma^{1}. 
\end{equation} 
The $\tau^{a}$ (a=1,3) are the usual Pauli matrices.
The supersymmetry transformation is given by
\begin{eqnarray}\label{susytransfn}
&&\delta\varphi=\bar{\epsilon}\psi(x),\nonumber\\
&&\delta\psi=\partial_{\mu}\varphi(x)\gamma^{\mu}\epsilon
+F(x)\epsilon=\delslash\varphi(x)\epsilon + F(x)\epsilon,
\nonumber\\
&&\delta F=\bar{\epsilon}\gamma^{\mu}\partial_{\mu}\psi
=\bar{\epsilon}\delslash\psi(x).
\end{eqnarray}
The supercharge which generates (\ref{susytransfn})
\begin{eqnarray}
&&\bar{\epsilon}Q\equiv 
\bar{\epsilon}_{\alpha} 
\frac{\partial}{\partial\bar{\theta}_{\alpha}}
-\bar{\epsilon}\gamma^{\mu}\theta\partial_{\mu}
\end{eqnarray}
and the covariant derivative 
\begin{eqnarray}
&&\bar{\eta}D\equiv \bar{\eta}_{\alpha} 
\frac{\partial}{\partial\bar{\theta}_{\alpha}}
+\bar{\eta}\gamma^{\mu}\theta\partial_{\mu}
\end{eqnarray}
commute with each other, $[\bar{\epsilon}Q,\bar{\eta}D]=0$.

The present model accommodates the kink solution which has been 
studied by many authors in the past~\cite{Dashen:1974cj, kaul}
with the conclusion reached by the majority of these authors
that supersymmetry
leads to vanishing corrections to mass and central charge. 
However, several years ago it was noted that the mass of the 
supersymmetric kink does receive quantum corrections, whereas the 
central charge (the integral of a total space derivative) did 
not seem to get corrected~\cite{Rebhan:1997iv,Nastase:1998sy}. This seemed to 
violate the BPS bound~\cite{R3}. It was therefore 
conjectured that the kink system contains a new 
anomaly~\cite{Nastase:1998sy}, 
and the existence of a central-charge anomaly was first established 
in~\cite{Shifman:1998zy}. It 
belongs to an anomaly multiplet of which the superconformal 
anomaly and the trace anomaly are also 
partners. One can also define a multiplet of 
conformal anomalies, and then the central charge itself is the 
anomaly in the conservation of a corresponding conformal 
central charge current~\cite{Rebhan:2002yw}. 

Recently a path integral approach to the anomalies in superspace
 was undertaken~\cite{fvn}, and a Ward identity in superspace 
was derived~\footnote{This Ward identity was derived by making 
a supersymmetry transformation with local ($x$-dependent and 
$\theta$-dependent) parameter. Because the action is only 
rigidly 
supersymmetric, the variation of the action produced the 
currents in (\ref{currwardid}), while the Jacobian yielded the anomalies (the 
last terms in (\ref{currwardid})).} in which the anomaly was due to the 
Jacobian of the 
path integral~\cite{fuji1,fuji2,konishi,shizuya}. Expanding 
this identity in terms of 
$\theta$, the following set of $x$-space Ward identities was 
obtained 
\begin{eqnarray}\label{currwardid}
&&(\gamma_{\mu}\tilde{J}^{\mu})(x)=(\gamma_{\mu}j^{\mu})(x)
-\frac{\hbar g}{\pi}\psi(x),\nonumber\\
&&\tilde{T}_{\mu}^{\ \mu}(x)=(T_{\mu}^{\ \mu})(x)
+\frac{\hbar g}{2\pi}F(x),\nonumber\\
&&\tilde{\zeta}^{\mu}(x) 
=(\zeta^{\mu})(x)
+\frac{\hbar g}{2\pi}
\epsilon^{\mu\sigma}\partial_{\sigma}\varphi(x),\nonumber\\
&&\partial_{\mu}j^{\mu}(x)=\frac{\hbar g}{2\pi}\delslash\psi(x).
 \end{eqnarray} These are operator relations, to be used 
inside the correlation functions.
They contain the supercurrent $j^{\mu}$ 
(and $\tilde{J}^{\mu}$), energy-momentum
tensor $T_{\mu\nu}$ (and $\tilde{T}_{\mu\nu}$) and central 
charge current $\zeta_{\mu}$ (and $\tilde{\zeta}_{\mu}$), 
respectively.
The full expressions of these operators read~\cite{fvn} 
\begin{eqnarray}\label{fullexpr}
j^{\mu}(x)&=&-[\delslash\varphi(x)
+U(\varphi(x))]\gamma^{\mu}
\psi(x),\nonumber\\
&&\nonumber\\ \tilde{J}^{\mu}(x)&=&
-[\delslash\varphi(x)-F(x)]\gamma^{\mu}
\psi(x)\nonumber\\
&=&j^{\mu}-\frac{\hbar g}{2\pi}\gamma^{\mu}\psi,\nonumber\\
T_{\mu\nu}(x)&=&
\partial_{\mu}\varphi\partial_{\nu}\varphi
-\frac{1}{2}\eta_{\mu\nu}[(\partial^{\rho}\varphi)
(\partial_{\rho}\varphi)-FU]\nonumber\\
&+&\frac{1}{4}\bar{\psi}[\gamma_{\mu}\partial_{\nu}+\gamma_{\nu}
\partial_{\mu}]\psi
-\frac{1}{4}\eta_{\mu\nu} 
[\bar{\psi}\gamma^{\rho}\partial_{\rho}\psi
+2g\varphi \bar{\psi}\psi],\nonumber\\
\tilde{T}_{\mu\nu}(x)&=& 
\partial_{\mu}\varphi\partial_{\nu}\varphi
-\frac{1}{2}\eta_{\mu\nu}[(\partial^{\rho}\varphi)
(\partial_{\rho}\varphi)+F^{2}]
+\frac{1}{4}\bar{\psi}[\gamma_{\mu}\partial_{\nu}+\gamma_{\nu}
\partial_{\mu}]\psi\nonumber\\
&=&T_{\mu\nu}(x)
+\eta_{\mu\nu}\frac{\hbar g}{4\pi}F(x),
\nonumber\\
&&\nonumber\\
\zeta_{\mu}(x)&=&\epsilon_{\mu\nu}\partial^{\nu}\varphi(x)
U(\varphi), \nonumber\\
&&\nonumber\\
\tilde{\zeta}_{\mu}(x)&=&-\epsilon_{\mu\nu}
\partial^{\nu}\varphi(x)F(x)
\nonumber\\
&=&\zeta_{\mu}(x)
+\frac{\hbar g}{2\pi}\epsilon_{\mu\nu}\partial^{\nu}\varphi(x)
\end{eqnarray}
with $U(\varphi)=g(\varphi^{2}(x)-v^{2}_{0})$ and 
$\epsilon_{01}=-1$. Another Ward identity 
in superspace was used to prove that $F+U$ may be replaced by 
$-\frac{\hbar g}{2\pi}$ inside composite operators.

It was shown in~\cite{fvn} that the currents $\tilde{J}^{\mu}$, 
$\tilde{T}_{\mu\nu}$ and $\tilde{\zeta}_{\mu}$ are conserved at 
the quantum level and, being $g$-independent, are free from 
explicit symmetry breaking terms  but contain anomalies.
On the other hand, the currents 
$j^{\mu}$, $T_{\mu\nu}$ and $\zeta_{\mu}$ are free from 
anomalies but are not conserved (except for the topological 
current $\zeta_{\mu}$) and contain explicit symmetry breaking 
terms as is clear from the presence of $g$-dependent terms.

\section{Local counter term and supersymmetry algebra}

It was shown in~\cite{fvn} that  
the following supersymmetry algebra is satisfied by the 
conserved supersymmetry charges 
$\tilde{Q}^{\alpha}=\int dx \tilde{J}^{0,\alpha}(x)$ 
\begin{eqnarray}\label{Qtildealgebra}
i\{\tilde{Q}^{\alpha},\tilde{Q}^{\beta}\}
&=&-2(\gamma^{\mu})^{\alpha\beta}\tilde{P}_{\mu}
-2\tilde{Z}(\gamma_{5})^{\alpha\beta}
\end{eqnarray}
where we defined 
\begin{eqnarray}
&&\tilde{P}_{\mu}=\int dx \tilde{T}_{0\mu}(x), 
\ \ \ \tilde{H}=\tilde{P}_{0} ,\nonumber\\
&&\tilde{Z}=\int dx \tilde{\zeta}_{0}(x)
=-\int dx\tilde{\zeta}^{0}(x).
\end{eqnarray}

It was also noted in~\cite{fvn} that the same supersymmetry 
algebra is satisfied  
by the naive (Noether) supersymmetry charges 
$Q^{\alpha}=\int dx j^{0,\alpha}(x)$ 
\begin{eqnarray}\label{Qalgebra} 
i\{Q^{\alpha},Q^{\beta}\} 
&=&-2(\gamma^{\mu})^{\alpha\beta}P_{\mu}
-2Z(\gamma_{5})^{\alpha\beta}
\end{eqnarray}
where we defined 
\begin{eqnarray}
&&P_{\mu}=\int dx T_{0\mu}(x), 
\ \ \ H=P_{0} ,\nonumber\\
&&Z=\int dx \zeta_{0}(x)
=-\int dx\zeta^{0}(x).
\end{eqnarray} 
The second algebra (\ref{Qalgebra}) was dismissed in 
\cite{fvn} because $j^{\mu,\alpha}(x)$ and $T_{\mu \nu}(x)$ 
are not conserved.  The BPS bound is then saturated 
in the algebra (\ref{Qtildealgebra}) by the trace and central 
charge 
anomalies\footnote{The trace of stress tensor is given by 
$\tilde{T}^{\mu}_{\ \mu}=-\tilde{T}_{00}+\tilde{T}_{11}
=-\tilde{T}_{00}$ since 
$\langle \tilde{T}_{11}\rangle=0$, and yields the total mass 
as a sum of regular and anomalous contributions.
For the supersymmetric kink, the regular quantum
corrections cancel if one uses minimal renormalization
with only a mass counter term such that
tadpoles vanish.}.

 What we would like to clarify in this note is the nature of
 the anomalous contributions appearing in (\ref{currwardid}). The superspace 
analysis of \cite{fvn} as summarized in the preceding chapter 
allows one to recognize that the 
the precise form of anomalies is modified if 
one adds the counter term
\begin{equation}\label{general-c}
{\cal L}_{counter}=-c\frac{\hbar g}{2\pi}\phi(x,\theta) 
\end{equation} to the original Lagrangian (\ref{Lagr}) where $c$ stands
 for a 
numerical constant. In particular, for $c=1$ the counter term  
\begin{equation}\label{Lc}
{\cal L}_{c}=-\frac{\hbar g}{2\pi}\phi(x,\theta)
\end{equation}
removes  all anomalies in (\ref{currwardid}) and 
(\ref{fullexpr}) proportional to $\hbar$.

The theory with extra local counter term is defined by 
\begin{equation}\label{Ltotal} 
{\cal L}_{total}={\cal L}+{\cal L}_{c} 
=\frac{1}{4}\bar{D}_{\alpha}\phi D^{\alpha} \phi
+\frac{1}{3}g\phi^{3}-gv^{2}_{0}\phi
-\frac{\hbar g}{2\pi}\phi(x,\theta).
\end{equation}
The {\em usual} counter term on the other hand is contained 
within $v_0^2=v^2+\delta v^2$, and we adopt the following 
minimal renormalization 
condition~\cite{Rebhan:1997iv,Rebhan:2002uk}
\begin{equation} 
U=g(\varphi^{2}-v^{2}_{0})=g(\varphi^{2}-v^{2}-\delta v^{2})
=g([\varphi^{2}]_{ren}-v^{2})
\end{equation}
at the one-loop level in the trivial vacuum; the counter term 
$\delta v^{2}$ is defined to remove the one-loop tadpole of 
$\varphi$  {\em prior to the addition of} ${\cal L}_{c}$.
Here $[\varphi^{2}]_{ren}$ stands for the 
renormalized operator which gives rise to a finite quantity 
when inserted into Green's functions.

The Lagrangian (\ref{Ltotal}) can be  obtained 
from the one without extra local counter term by the 
replacement $v^{2}_{0}\rightarrow v^{2}_{0}+\frac{\hbar}{2\pi}$.
 It is shown that both of the above supersymmetry algebras
(\ref{Qtildealgebra}) and (\ref{Qalgebra}) 
hold in the theory with the extra local counter term
if one makes the above replacement of the vacuum parameter. 
This is intuitively understood by noting that the canonical 
equal-time commutators are independent of the vacuum parameter. 
In the theory with extra local counter term, the conserved 
quantities are given by (using the notation in (\ref{fullexpr}))
\begin{eqnarray}
\tilde{J}^{\mu,\alpha}(x)
_{v^{2}_{0}\rightarrow v^{2}_{0}+\frac{\hbar}{2\pi}},\ \ 
\tilde{T}_{\mu\nu}(x)
_{v^{2}_{0}\rightarrow v^{2}_{0}+\frac{\hbar}{2\pi}},\ \ 
\tilde{\zeta}_{\mu}(x)
_{v^{2}_{0}\rightarrow v^{2}_{0}+\frac{\hbar}{2\pi}} 
\end{eqnarray} 
but these operators are expressed in terms of the operators 
in (\ref{fullexpr}) for the theory {\em without} the extra counter term as 
follows  
\begin{eqnarray}\label{shifts}
&&\tilde{J}^{\mu,\alpha}(x)
_{v^{2}_{0}\rightarrow v^{2}_{0}+\frac{\hbar}{2\pi}}
=j^{\mu}
_{v^{2}_{0}\rightarrow v^{2}_{0}+\frac{\hbar}{2\pi}} 
-\frac{\hbar g}{2\pi}\gamma^{\mu}\psi=j^{\mu}_{v^{2}_{0}},
\nonumber\\
&&\tilde{T}_{\mu\nu}(x)
_{v^{2}_{0}\rightarrow v^{2}_{0}+\frac{\hbar}{2\pi}}
=T_{\mu\nu}(x)
_{v^{2}_{0}\rightarrow v^{2}_{0}+\frac{\hbar}{2\pi}}
+\eta_{\mu\nu}\frac{\hbar g}{4\pi}F(x)=T_{\mu\nu}(x)_{v^{2}_{0}}
,\nonumber\\
&&\tilde{\zeta}_{\mu}(x)
_{v^{2}_{0}\rightarrow v^{2}_{0}+\frac{\hbar}{2\pi}}
=\zeta_{\mu}(x)
_{v^{2}_{0}\rightarrow v^{2}_{0}+\frac{\hbar}{2\pi}}
+\frac{\hbar g}{2\pi}\epsilon_{\mu\nu}\partial^{\nu}\varphi(x)
=\zeta_{\mu}(x)_{v^{2}_{0}}.
\end{eqnarray}
We thus have 
\begin{equation}\label{Ttildeshifted}
\tilde{T}_{\mu\nu}(x)
_{v^{2}_{0}\rightarrow v^{2}_{0}+\frac{\hbar}{2\pi}}
= T_{\mu\nu}(x)_{v^{2}_{0}}. 
\end{equation}

We note that  the conservation conditions of 
$\tilde{T}_{\mu\nu}(x)
_{v^{2}_{0}}$ and $T_{\mu\nu}(x)_{v^{2}_{0}}$ are expressed in 
the path integral formulation as
\begin{eqnarray}
\langle \partial^{\mu}\tilde{T}_{\mu\nu}(x)
_{v^{2}_{0}}\rangle 
=\int{\cal D}\phi \partial^{\mu}\tilde{T}_{\mu\nu}(x)
_{v^{2}_{0}}\exp\{i\int dx {\cal L} \}=0
\end{eqnarray}
and we obtain, replacing $v_{0}^{2}$ by 
$v^{2}_{0}+\frac{\hbar}{2\pi}$ (which replaces ${\cal L}$ by 
${\cal L}+{\cal L}_{c}$), and using (\ref{Ttildeshifted}) 
\begin{eqnarray}
\langle \partial^{\mu}T_{\mu\nu}(x)
_{v^{2}_{0}}\rangle 
=\int{\cal D}\phi \partial^{\mu}T_{\mu\nu}(x) _{v^{2}_{0}}
\exp\{i\int dx[{\cal L}+{\cal L}_{c}] \}=0. 
\end{eqnarray} 
The operator $T_{\mu\nu}(x)_{v^{2}_{0}}$, which is not conserved 
for the theory defined by ${\cal L}$, becomes conserved for the 
theory specified by ${\cal L}+{\cal L}_{c}$; different equations
  of motion give rise to different conservation properties.

As for the trace anomaly, we note that  
\begin{eqnarray}
T_{\mu}^{\ \mu}(x)_{v^{2}_{0}} &=&[gF(\varphi^{2}-v^{2}_{0})
-g\varphi\bar{\psi}\psi]\nonumber\\
&=&[gF(\varphi^{2}-v^{2}_{0})-g\varphi\bar{\psi}\psi]
_{v^{2}_{0}\rightarrow v^{2}_{0}+\frac{\hbar}{2\pi}}
+\frac{\hbar g}{2\pi}F(x)\nonumber\\
&=&T_{\mu}^{\ \mu}(x)
_{v^{2}_{0}\rightarrow v^{2}_{0}+\frac{\hbar}{2\pi}}
+\frac{\hbar g}{2\pi}F(x)
\end{eqnarray}
and thus the trace of the conserved tensor 
$T_{\mu\nu}(x)_{v^{2}_{0}}$ in the theory with the counter term 
in fact 
contains the trace anomaly in addition to 
$T_{\mu}^{\ \mu}(x) _{v^{2}_{0}\rightarrow v^{2}_{0}
+\frac{\hbar}{2\pi}}$ which 
is the explicit breaking term for the theory with the counter 
term. 
In connection 
with this observation, we emphasize that
if one makes the replacement 
$v^{2}_{0}\rightarrow v^{2}_{0}+\frac{\hbar}{2\pi}$ for the 
combination of the explicit symmetry breaking term and the trace
anomaly for the theory {\em without} the counter term,  
the cancellation of the trace anomaly by a variation of the 
explicit symmetry breaking term takes place regardless of 
regularization schemes. Our analysis is thus not specific to 
the path integral formulation.

Similarly, for the $\gamma$-trace of the central charge current 
(which is equivalent to the current itself) we have the anomaly 
relation 
\begin{eqnarray} \zeta_{\mu}(x)_{v^{2}_{0}}&=&
\zeta_{\mu}(x)
_{v^{2}_{0}\rightarrow v^{2}_{0}+\frac{\hbar}{2\pi}}
+\frac{\hbar g}{2\pi}\epsilon_{\mu\nu}\partial^{\nu}\varphi(x)\,.
\end{eqnarray}

In the theory with  extra local counter term, 
the relevant conserved
quantities are thus given by the operators appearing on the 
right-hand sides of (\ref{shifts}). The supersymmetry algebra 
satisfied by the conserved quantities is then given by the 
naive algebra (\ref{Qalgebra}). The BPS bound in the theory 
with the extra local counter term
is thus formally satisfied without anomaly contributions.

However, we note some differences between the local counter term 
in the present context and  counter terms in ordinary gauge 
theories. For example, the main physics issue in the present 
context is the saturation of the BPS bound, which is saturated 
in both 
cases with or without extra local counter term, though the 
actual 
manner of saturation differs in these two cases with or 
without the extra local counter term. 
Also, the inevitable 
trace anomaly for the conserved energy-momentum tensor, 
which has been established by a variety of 
ways~\cite{Nastase:1998sy}$\sim$\cite{Rebhan:2002yw}, 
is present 
in the theory with or without the extra local 
counter term. 

In the present problem, the kink mass is another important 
physical quantity, and we want to comment on  how it remains 
invariant under the addition of extra local counter terms. 
The fact that the counter term (\ref{Lc}) removes all the 
anomalies suggests that all the anomalies are generated if one 
adds an extra local term
\begin{equation}
{\cal L}_{anomaly}=\frac{\hbar g}{2\pi}\phi(x,\theta) 
\end{equation} to the Lagrangian, by pretending as if the path 
integral measure generates no non-trivial Jacobians. The 
term ${\cal L}_{anomaly}$ plays a role of a  
``Wess-Zumino term''. This picture is convenient when one 
analyzes the effects of the anomalies in the framework of 
the effective Lagrangian and the effective potential.
 
In the above picture with ${\cal L}_{anomaly}$, the original 
theory is effectively described by 
\begin{equation}
{\cal L}_{eff}={\cal L}+{\cal L}_{anomaly} 
=\frac{1}{4}\bar{D}_{\alpha}\phi D^{\alpha} \phi
+\frac{1}{3}g\phi^{3}-gv^{2}_{0}\phi
+\frac{\hbar g}{2\pi}\phi(x,\theta)
\end{equation}
combined with the {\em naive} path integral measure. 
The net effect of the anomalies is thus the replacement of 
the vacuum value 
$v^{2}_{0}\rightarrow v^{2}_{0}- \frac{\hbar}{2\pi}$, or after 
the renormalization
\begin{equation}
v^{2}\rightarrow v^{2}- \frac{\hbar}{2\pi}.
\end{equation}
The justification of this picture is given when one starts with 
the naive operators
\begin{eqnarray} j^{\mu}(x)&=&-[\delslash\varphi(x)
+U(\varphi(x))]\gamma^{\mu}
\psi(x),\nonumber\\
T_{\mu\nu}(x)&=&
\partial_{\mu}\varphi\partial_{\nu}\varphi
-\frac{1}{2}\eta_{\mu\nu}[(\partial^{\rho}\varphi)
(\partial_{\rho}\varphi)-FU]\nonumber\\
&+&\frac{1}{4}\bar{\psi}[\gamma_{\mu}\partial_{\nu}+\gamma_{\nu}
\partial_{\mu}]\psi
-\frac{1}{4}\eta_{\mu\nu} 
[\bar{\psi}\gamma^{\rho}\partial_{\rho}\psi
+2g\varphi \bar{\psi}\psi],\nonumber\\
\zeta_{\mu}(x)&=&\epsilon_{\mu\nu}\partial^{\nu}\varphi(x)
U(\varphi), 
\end{eqnarray}
with $U(\varphi)=g(\varphi^{2}(x)-v^{2}_{0})$ and 
$\epsilon_{01}=-1$. When one makes the above replacement 
$v^{2}_{0}\rightarrow v^{2}_{0}- \frac{\hbar}{2\pi}$ in these 
naive operators, all the operators with tilde in (\ref{fullexpr}), which 
give the correct anomalies, are generated.

The elementary fermion or boson mass parameter in the 
asymptotic region $m=2gv$, which is not the pole mass itself but 
nevertheless defines a
physical parameter, is thus 
given by the above replacement as
\begin{equation}
m=2g\sqrt{v^{2}-\frac{\hbar}{2\pi}}.
\end{equation}
On the other hand, it was shown in~\cite{fvn} that the total 
central charge, which is 
equal to the total energy of the kink vacuum in the theory 
without the extra local counter term,  is given by 
\begin{equation}
M=\frac{4}{3}g(v^{2}-\frac{\hbar}{2\pi})^{3/2}
\end{equation}
where $-\frac{\hbar}{2\pi}$ arises from the effect of the 
central charge anomaly. This kink mass is written in terms of 
the modified fermion mass parameter as
\begin{equation}\label{Mmod}
M=\frac{1}{6g^{2}}m^{3}
\end{equation}
which has the same form as the naive kink mass expressed in 
terms of the naive fermion or boson mass parameter, as is the 
case in the theory with the extra local counter term 
which removes 
all the anomalies. We thus have the same physical relation 
between physical masses with 
or without the extra local counter term.

 It should be noted, however, that the transition from the 
original Lagrangian to the one given by 
(\ref{Ltotal}) corresponds to a modification of the 
renormalization scheme in the present model. With 
(\ref{Ltotal}) the condition which determines $\delta v^2$ is 
in effect one where the tadpole diagrams are renormalized 
precisely such as to cancel additional terms in 
(\ref{Mmod}).\footnote{If one would like to have a renormalization
scheme 
where all tadpoles vanish even after the transition to 
(\ref{Ltotal}), this can be achieved by allowing for a finite 
renormalization of $g$ such that (\ref{Mmod}) is maintained 
(see Ref.~\cite{Rebhan:2002uk} for a detailed discussion
of possible renormalization schemes in this model). }
Thus physical quantities as well as the property of BPS 
saturation are left intact simply because a change of 
renormalization prescriptions does not change physics; yet, on 
a formal level the anomalous contributions to both sides of the 
BPS bound are modified.

\section{Discussion and conclusion}

We have shown that the extra local counter term (\ref{Lc}) 
formally removes 
all the anomalies from the supersymmetry algebra in the model (\ref{Lagr}).
We have also shown that the
physical consequences are not modified by the counter term. 
It is possible to show this equivalence for a more general 
choice of the counter term. At 
the same 
time we explained that the trace of the conserved 
energy-momentum tensor consists of the explicit symmetry 
breaking terms 
plus a unique trace anomaly regardless of counter terms, and 
similarly the $\gamma$-trace of the central charge current. 

In the context of the present problem  we thus conclude that  
the 
issue whether the anomalies are spurious or not is immaterial. 
Rather, we have seen that different descriptions based 
on  different bare Lagrangians give rise to the same 
physical consequences. In particular, the original description 
with equal anomalies in the trace and central charge provides a 
consistent and well-defined description of the quantized 
supersymmetric kink model. This description is quite informative
 with respect to physics, such as the inevitable 
appearance of a trace anomaly for the conserved energy-momentum 
tensor. The 
quantum modification of the central charge, as manifested as 
the central charge anomaly, may take place for a more general 
class of Abelian topological quantities.  On the other hand,  
the description with the extra counter term which removes 
all the 
anomalies is less 
informative though the physical content of the model is 
succinctly expressed.\footnote{The continuous number of ways
to describe the same physics, which is parametrized by the 
counter term (\ref{general-c}), is to some extent analogous to the 
$R_{\xi}$-gauge formulation of the Higgs mechanism. The unitary 
gauge 
$\xi=0$ succinctly expresses the physical content, but other 
gauges are more informative about the essence 
of the Higgs mechanism. The  would-be 
Nambu-Goldstone bosons appear in general gauge conditions 
though they disappear in the unitary gauge. The question if the 
Nambu-Goldstone bosons are ``spurious'' or not in the Higgs 
mechanism may be compared to the question if the anomalies 
are ``spurious'' or not in the analysis of BPS bound in the 
present kink model. One may say that the anomalies are  
essential 
to understand the quantized supersymmetric kink 
properly, even if they do not explicitly appear in the  
description with $c=1$, just like the Nambu-Goldstone
bosons in the Higgs mechanism.}
Another way to 
understand why the extra local counter term, which formally 
removes explicit anomalies from the supersymmetry 
algebra, does not
change physics, is to recognize, as we have discussed,
that it corresponds to a change of the 
renormalization scheme, and this does not change physics. 
Only the formal analysis of how the BPS bound 
is saturated is changed.

This is in sharp contrast to the case in gauge
theory where the description with counter terms which eliminate 
all the (spurious) gauge anomalies is the {\em unique} 
description of the physical theory.
\\

\noindent {\bf Acknowledgments}
\\

One of us (KF) thanks all the members of the C.N.Yang Institute 
for Theoretical Physics at Stony Brook for their hospitality.


\begin{thebibliography}{99}


\bibitem{R3}
E. Witten, D. Olive, Phys. Lett. {\bf 78B} (1978) 97.

\bibitem{Dashen:1974cj}
R.~Dashen, B.~Hasslacher, A.~Neveu, Phys. Rev. {\bf D10} (1974) 
4130.
\\
J.-L. Gervais, A.~Neveu~(eds.), Phys. Rept. {\bf 23} (1976) 237.
 \\ 
L.~D. Faddeev, V.~E. Korepin, Phys. Rept. {\bf 42} (1978) 1. \\ 
J.~F. Schonfeld, Nucl. Phys. {\bf B161} (1979) 125. 

\bibitem{kaul} R.~K. Kaul, R.~Rajaraman, Phys. Lett. {\bf B131}
 (1983) 357. \\ 
C.~Imbimbo, S.~Mukhi, Nucl. Phys. {\bf B247} 
(1984) 471. \\ 
H.~Yamagishi, Phys. Lett. {\bf B147} (1984) 425.
 \\ 
A.~Chatterjee, P.~Majumdar, Phys. Rev. {\bf D30} (1984) 844;
 Phys. Lett. {\bf B159} (1985) 37. \\ 
A.~Uchiyama, Prog. Theor. Phys. {\bf 75} (1986) 1214;
Nucl. Phys. {\bf B278} (1986) 121. \\ 
L.~J. Boya, J.~Casahorr{\'a}n, J. Phys. {\bf A23} (1990) 1645.

\bibitem{Rebhan:1997iv}
A.~Rebhan, P.~van Nieuwenhuizen, Nucl. Phys. {\bf B508} (1997) 
449.

\bibitem{Nastase:1998sy}
H.~Nastase, M.~Stephanov, P.~van Nieuwenhuizen, A.~Rebhan, 
Nucl. Phys. {\bf B542} (1999) 471.

\bibitem{Goldhaber:2000ab}
A.~S. Goldhaber, A.~Litvintsev, P.~van Nieuwenhuizen, Phys. Rev. {\bf D64} 
(2001)
  045013;
A.~S. Goldhaber, A.~Rebhan, P.~van Nieuwenhuizen, R.~Wimmer, Phys. Rev. 
{\bf D66}
  (2002) 085010.

\bibitem{Rebhan:2002uk}
A.~Rebhan, P.~van Nieuwenhuizen, R.~Wimmer, New J. Phys. {\bf 4}
 (2002) 31.

\bibitem{Graham:1998qq}
N.~Graham, R.~L. Jaffe, Nucl. Phys. {\bf B544} (1999) 432.

\bibitem{Shifman:1998zy}
M.~A. Shifman, A.~I. Vainshtein, M.~B. Voloshin, Phys. Rev. 
{\bf D59} (1999) 045016.

\bibitem{losev}
A.~Losev, M.~A. Shifman, A.~I. Vainshtein, New J. Phys. {\bf 4} (2002)
  21.

\bibitem{Bordag:2002dg}
M.~Bordag, A.~S. Goldhaber, P.~van Nieuwenhuizen, 
D.~Vassilevich, Phys. Rev. {\bf D66} (2002) 125014.

\bibitem{GLN}
A.~S. Goldhaber, A.~Litvintsev, P.~van Nieuwenhuizen, 
Phys. Rev. {\bf D67} (2003) 105021.

\bibitem{Rebhan:2002yw}
A.~Rebhan, P.~van Nieuwenhuizen, R.~Wimmer, Nucl. Phys. 
{\bf B648} (2003) 174.

\bibitem{fvn}
K. Fujikawa,  P. van Nieuwenhuizen, \textit{
Topological anomalies from the path integral measure in 
superspace}, hep-th/0305144, Ann. of Phys. (in press). 

\bibitem{fuji1}
K. Fujikawa, Phys. Rev. Lett. {\bf 42} (1979) 1195; Phys. Rev. 
{\bf D21} (1980) 2848.

\bibitem{fuji2}
K. Fujikawa, Phys. Rev. Lett. {\bf 44} (1980) 1733; Phys. Rev. 
{\bf D23} (1981) 2262.

\bibitem{konishi}
K. Konishi, K. Shizuya, Nuovo Cim. {\bf 90A} (1985) 111.

\bibitem{shizuya}
K. Shizuya, Phys. Rev. {\bf D35} (1987) 1848; {\bf D35} (1987) 
2550.

\end{thebibliography}
\end{document}